\documentclass[a4,twocolumn]{revtex4}
\usepackage{graphicx}
\usepackage{amssymb,amsfonts,amsmath,mathrsfs,gensymb}
\usepackage{color}
\usepackage{multirow}
\usepackage[normalem]{ulem}

\begin{document}
\title{ATP dependent NS3 helicase interaction with RNA: insights from molecular simulations}
\author{%
Andrea P\'erez-Villa\,$^{1}$,
Maria Darvas\,$^{1}$,
Giovanni Bussi \,$^{1,}$%
\footnote{To whom correspondence should be addressed.
Email: bussi@sissa.it
}
}

\address{%
$^{1}$
Scuola Internazionale Superiore di Studi Avanzati, International School for Advanced Studies,
265, Via Bonomea I-34136 Trieste, Italy
}

\begin{abstract}
Non structural protein 3 (NS3) helicase from hepatitis C virus is an enzyme that unwinds and translocates along nucleic acids
with an ATP-dependent mechanism and has a key role in the replication of the viral RNA.
An inchworm-like mechanism for translocation has been proposed based on crystal
structures and single molecule experiments. We here perform atomistic molecular dynamics in explicit solvent on the microsecond
time scale of the available experimental structures. We also construct and simulate
putative intermediates for the translocation process, and we perform non-equilibrium
targeted simulations to estimate their relative stability.
For each of the simulated structures we carefully characterize the available conformational space,
the ligand binding pocket, and the RNA binding cleft.
The analysis of the hydrogen bond network and of the non-equilibrium trajectories
indicates an ATP-dependent stabilization of one of the protein conformers.
Additionally, enthalpy calculations suggest that entropic effects might be crucial for the stabilization of the experimentally observed structures.

\end{abstract}
\maketitle

\section{Introduction}
Non-structural protein 3 (NS3)
is a molecular motor encoded by Hepatitis C virus (HCV) consisting of a N-terminal region with a serine protease domain
and a C-terminal superfamily 2 (SF2) helicase that unwinds and translocates on
nucleic acids. This protein translocates on single-stranded ribonucleic acid (ssRNA)
in the 3'$\rightarrow$5' direction through a periodic stepwise mechanism \cite{emboj_pang, annurev_pyle, pyle-met-enz, colizzi2012rna}.
Translocation is ATP dependent, so that this enzyme
has been also classified as a DExH box ATPase \cite{schwer}.
Understanding the mechanism of action of this molecular motor
at the atomistic level is fundamental since
the NS3 helicase domain has been proposed as a target
for the development of a antiviral agents \cite{raney2010}.

NS3 helicase has been characterized by single-molecule experiments \cite{myong, arunajadai, bustamante, dumont, dumont2}
and biochemical essays \cite{fluorometric,frick2012, ventura2014}.
Although sometime acting as a dimer or as an oligomer \cite{jennings2009, raney2008},
NS3 also functions as a monomer, similarly to other SF1 and SF2 helicases \cite{annurev_wigley,rev_jankowsky}.
The N-terminal protease domain affects the binding of NS3
to RNA and plays an important role for the reaction kinetics \cite{patel2010}.
However, the protease domain is not essential for the helicase activity \cite{gu_rice,pyle_jbc2007},
thus the helicase domain (NS3h) can be characterized in isolation.
Interestingly, optical tweezer experiments have provided estimates of the number of
substeps per cycle, up to a resolution of single base pair \cite{arunajadai, bustamante}.

Fluorescence resonance energy transfer (FRET) \cite{myong} on the NS3-DNA complex suggested
a step of 3 bp with 3 hidden substeps where 1 bp is unwound per 1 ATP molecule
consumed following an inchworm mechanism. 
However, although single molecule experiments allow the kinetics of the mechanism to be captured,
they cannot provide detailed structural information. Additionally, the force applied during mechanical manipulation
is often much larger than the actual force felt by biopolymers \emph{in vivo} \cite{force-outeq1,force-outeq2}.
On the other hand, X-ray crystallography can provide detailed snapshots at atomistic resolution.
Only a few intermediate snapshots have been reported so far related to NS3h translocation on RNA \cite{yao1999, gu_rice, pyle-jmb},
and conformational differences between these snapshots have been interpreted using elastic network models
\cite{flechsig, mustafa}. In this context, molecular dynamics simulations \cite{tuckerman-book} with accurate force fields
could add dynamical information to the available crystal structures
providing a new perspective on the mechanism of action of this important molecular motor.

In this paper, we describe atomistic molecular dynamics (MD) simulations in explicit
solvent of NS3-ssRNA complex in the absence (apo) and presence of ATP/ADP. In order to 
understand the stability of the intermediates along the translocation cycle
we constructed putative intermediate structures. %
We used a recent version of the AMBER force field and performed
microsecond time scale simulations so as to provide statistically meaningful results.
Results are complemented with non-equilibrium targeted molecular dynamics so as to
assess the relative stability of the apo, ADP, and ATP structures.
Experimentally determined structures are shown to be stable within this timescale.
Both the experimental structures and the putative intermediate ones are analyzed in details.
Only a handful of MD simulations have been reported on nucleic-acid/helicase complexes so far \cite{mustafa, schulten},
all of them on a much shorter time scale.
To the best of our knowledge, only a few MD simulations have been performed on the
microsecond timescale for
RNA-protein complexes of comparable or larger size \cite{bock2013energy, whitford2013connecting, krepl2015can},
and thus our results provide a valuable benchmark for state-of-the-art molecular dynamics of
these systems.

\section{Methods}

\begin{figure}\centering
 \includegraphics[scale=0.22]{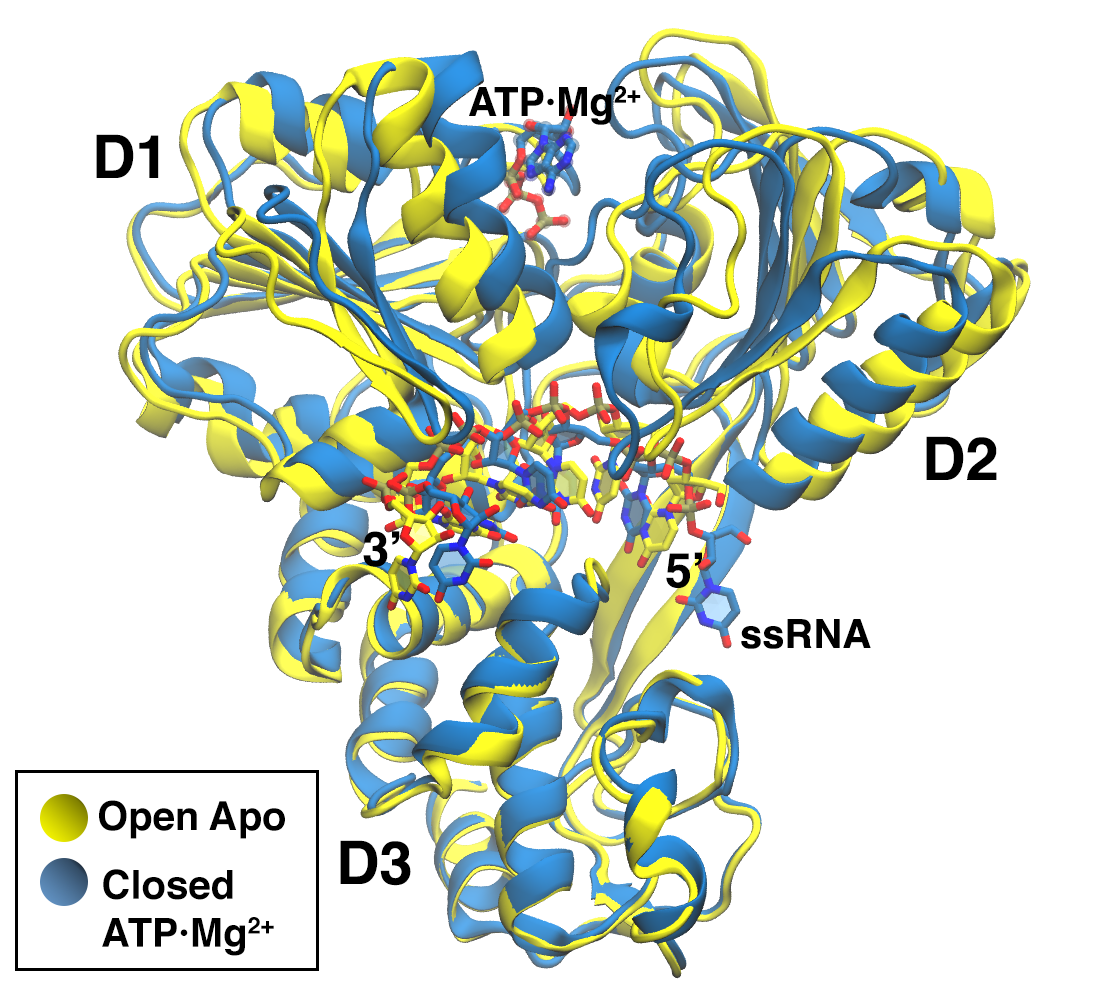}
 \caption{
Comparison between open-apo and closed-ATP crystal structures. 
NS3h is represented with ribbons, while ATP and ssRNA are displayed with sticks.
The gap between domain 1 (D1) and domain 2 (D2) is larger for the open conformation. 
The distance between centers of mass of D1 and D2 is $\sim$29 \AA\ in the open structure and $\sim$26 \AA\ in the closed one.
Water and ions are not shown.
}
 \label{fig:peptide}
\end{figure}

Atomistic molecular dynamics simulations in explicit solvent were performed for the NS3h-ssRNA complex
for three different molecular systems (apo, with ADP and with ATP)
starting from two different conformations (open and closed), for a total of six simulations.
In the open form the distance between domains D1 and D2 is larger than in the closed form.
Crystal structures are available in the protein data bank (PDB)
for the apo open form (PDB: 3O8C) and for the holo (ATP) closed form (PDB:3O8R)~\cite{pyle-jmb}.
The four missing combinations, namely open-ATP, open-ADP, closed-ADP, and closed-apo, were constructed as described below.
A superimposed representation of open-apo and closed-ATP crystal structures can be seen in Figure~\ref{fig:peptide}.
The molecular visualizations were generated using VMD \cite{vmd, tachyon}.

\subsection{Modeling the intermediate states}
The missing intermediate states were built based on other experimental
structures available from the PDB.
The closed-apo form was constructed by ATP removal from the available closed-ATP structure (PDB: 3O8R)~\cite{pyle-jmb}.
The open-ATP/ADP form was constructed by means of a structural alignment procedure,
adding the ATP/ADP to the available open-apo structure (PDB: 3O8C)~\cite{pyle-jmb}.
To properly place the ATP, we used the ATP coordinates from the closed-ATP structure (PDB: 3O8R)
after structural alignment of the binding pocket site, Motif I (Walker A, residues 204 to 211).
Due to the lack of a crystal structure for the NS3h HCV-ssRNA-ADP complex, other NS3 helicases of flaviviridae virus
were analyzed to characterize the ADP binding on the protein. The possible candidates that belong to this family of viruses are Dengue and Yellow
Fever proteins. Crystal structures of complex protein-ssRNA-ATP/ADP are available for Dengue virus (PDB: 2JLV and 2JLZ respectively)~\cite{dengue}.
Walker A is a well conserved motif across SF2 NS3 helicases, and a structural alignment showed that ATP and ADP placement is
virtually identical in PDB: 2JLV and PDB: 2JLZ of Dengue virus. We thus used coordinates of ADP taken from
PDB: 3O8R in the closed NS3h, replacing ATP, and in the open NS3h (PDB: 3O8C).

We observe that in reference \cite{mustafa} the open-ATP form was constructed by performing a targeted MD starting from the closed-apo structure.
We preferred here to use the structural alignment procedure discussed above to avoid potential artifacts resulting from the
non-equilibrium pulling in the generation of the starting points for our long MD simulations.

\subsection{Molecular dynamics simulations}
Each of the starting structures consisting of NS3 peptide (436 aminoacids),
polyUracil ssRNA (6 nucleotides), and, when present, ATP/ADP$\cdot$Mg$^{2+}$, was solvated in a box containing
31058 water molecules,
70 Na$^+$ ions, corresponding to a 0.1 M concentration,
and the missing Cl$^-$ ions to neutralize the total charge,
for a total of $\sim$100000 atoms (see Table SD1 for details).
GROMACS 4.6 program \cite{gromacs} with AMBER99sb*ILDN-parmbsc0-$\chi_{OL3}$ + AMBER99ATP/ADP force fields
were used.
This force field is based on AMBER99 force field~\cite{wang2000well}, and additionally implements
corrections to protein backbone~\cite{hornak2006comparison,amber_hummer},
protein side-chain \cite{amber_ildn},
RNA backbone \cite{amber_rna}, and glycosidic torsions \cite{zgarbov2011}
as well as parameters for ATP and ADP \cite{amber_atp} and Mg$^{2+}$ \cite{ff_villamg}.
Water molecules were described by the TIP3P model~\cite{tip3p}.
All the molecular dynamics simulations were performed on the isothermal-isobaric ensemble,
using the stochastic velocity rescaling thermostat at 300 K~\cite{bussi2007canonical}  and the Berendsen barostat
with an isotropic pressure coupling of 1 bar ~\cite{berendsen_p}.
The systems were simulated for 1$\mu$s each, with a 2 fs time step, using periodic boundary conditions,  the LINCS algorithm \cite{lincs} to constrain bonds, and the particle-mesh Ewald method~\cite{pme,smooth_pme} to account for long-range electrostatics.
The resulting setup for the apo simulations is thus similar to the one used in \cite{ferrarotti}.
The conformations obtained after 200 ns were extracted and used with randomized velocities as starting points for further
control simulations of 200 ns length each. The root-mean-square deviations after structural alignment (RMSD) \cite{kabsch}
of the long MD and of the control simulations were used to verify the stability of the simulated systems.

\subsection{Monitoring conformational changes}
To monitor structural fluctuations we used two order parameters based on a combination of 
RMSDs from the crystal structures.
In reference~\cite{mustafa} the difference between squared RMSD from open and closed structure was used to monitor
the conformational change. We here used a similar procedure
but we considered a subset of atoms from the peptide and the ssRNA that are mainly involved in the conformational change.
The full list of atoms is provided in Supplementary Data (Figure SD1 and Table SD2).
We observe that the difference between squared RMSDs is closely related to the progression
variable $S$ used in the context of path collective variables \cite{branduardi2007, ensing_PRL}.
We thus also introduce a variable $Z$ that measures the distance from a
hypothetical transition path obtained as a linear interpolation between the two experimental structures.
$S$ and $Z$ are thus defined as $S=\frac{R^2_{o}-R^2_{c}}{2R_{oc}}$
and $Z=\frac{R^2_{o}+R^2_{c}}{2}-S^2-\frac{R^2_{oc}}{4}$
Here $R_o$ and $R_c$ are the RMSDs from the open (PDB: 3O8C) and closed (PDB: 3O8R) structures respectively,
and $R_{oc}$ is the RMSD between the open and the closed structure.

Additionally, we performed a principal components analysis (PCA) using the same set of atoms \cite{pca}. The PCA was made
based on a single trajectory obtained by concatenating all the 6 simulations. Individual simulations were then projected on the
eigenvectors corresponding to the two largest eigenvalues.

\subsection{Hydrogen-bond analysis}
Hydrogen bonds are computed based on distance-angle geometric criteria \cite{hbonds}.
The cut-off radius for the distance donor-acceptor is of 3.5 \AA\ and the cut-off angle between acceptor-donor-hydrogen (0$\degree$ is the strongest interaction) is 30$\degree$. The \emph{g\_hbond} tool of GROMACS 4.6 program was used for the calculations, and the standard errors were estimated through a binning analysis.

\subsection{Stacking interactions}
In order to monitor stacking interactions between nucleobases and
between aromatic aminoacids and nucleobases we used two different procedures.
Intra RNA stacking was annotated by the baRNAba tool~\cite{barnaba}, which assigns 
one oriented bead to  each nucleotide to describe RNA structural properties.
For stacking between aromatic aminoacids and nucleobases we used a geometric criterion
where the two residues were considered as stacked if the distance between their center
of mass was less than 5 \AA{} and the angle between the two planes was less than 30$\degree$.
This latter analysis has been performed for selected interactions
that are observed in the crystal structure, namely Y241 with ligand adenine and for W501 with U7.
Distances and angles were computed with PLUMED~\cite{tribello2014plumed}.

\subsection{Electrostatic interactions}
Electrostatic interactions between protein-RNA and protein-ligand were monitored by computing the
Debye-H\"uckel interaction energy \cite{dh_book}. The calculation was made using PLUMED~\cite{tribello2014plumed}
with the implementation described in ref~\cite{trang}. An ionic strength of 0.1M was selected that corresponds to
a screening length of $\approx$ 10 \AA.

\subsection{Total enthalpy calculations}
Enthalpy is defined in the isothermal-isobaric ensemble as
$ H=\langle U\rangle+p\langle V\rangle$. Here $U$ is the potential energy, $p$ the external pressure, and $V$ the volume of the simulated box.
The enthalpy values were computed here with the \emph{g\_energy} tool of GROMACS 4.6 program.
For energy differences to be meaningful, simulations corresponding to different conformations (open or closed) of the same system
were prepared so as to contain exactly the same number and types of atoms.
Since total energy can be affected by
numerical details, all simulations were performed using identical settings on identical computers.
The first 200 ns of MD were performed using GPUs+CPUs, and the remaining 800 ns using CPUs only.
We obseved no significant difference between results obtained with GPUs+CPUs and results obtained with CPUs only.
We also recall that total energy calculation can be affected by statistical errors mostly due to fluctuations of solvent contributions.
We here computed errors using a binning analysis, and run simulations long enough for this error to be $<5$ kcal/mol.

\subsection{Targeted molecular dynamics}

To estimate the relative stability of the closed and open conformations for the apo/ADP/ATP structures, we performed short
targeted MD starting from the closed structures \cite{tmd} following a protocol similar to the one used in ref \cite{mustafa}. Namely, we applied a time dependent
harmonic restraint of stiffness 500 kcal/(mol \AA$^2$) to
the RMSD from the corresponding  open structures. RMSD was computed using all heavy atoms of protein and RNA.
The center of the restraint was moved from a given initial value to  zero during a 20 ns long simulation. During the first ns
the center was left at its original value, whereas during the following 19 ns the center was moved linearly in time to zero.
To avoid any bias,
we chose the initial value to be  equal to RMSD between the closed structure after equilibration and the open crystal in all the three cases,
and we repeated every simulation three times using a different random seed. This resulted in a total
of 27 independent simulations, corresponding to 3 different systems (apo/ADP/ATP), 3 different steering protocols
(starting from the initial RMSD of apo/ADP/ATP structures) and 3 different random seeds.
The relative stability of the closed and open structures was estimated by measuring the work performed during the pulling.

\section{Results}

\subsection{Structural analysis}
\begin{figure}\centering
 \includegraphics[width=\columnwidth]{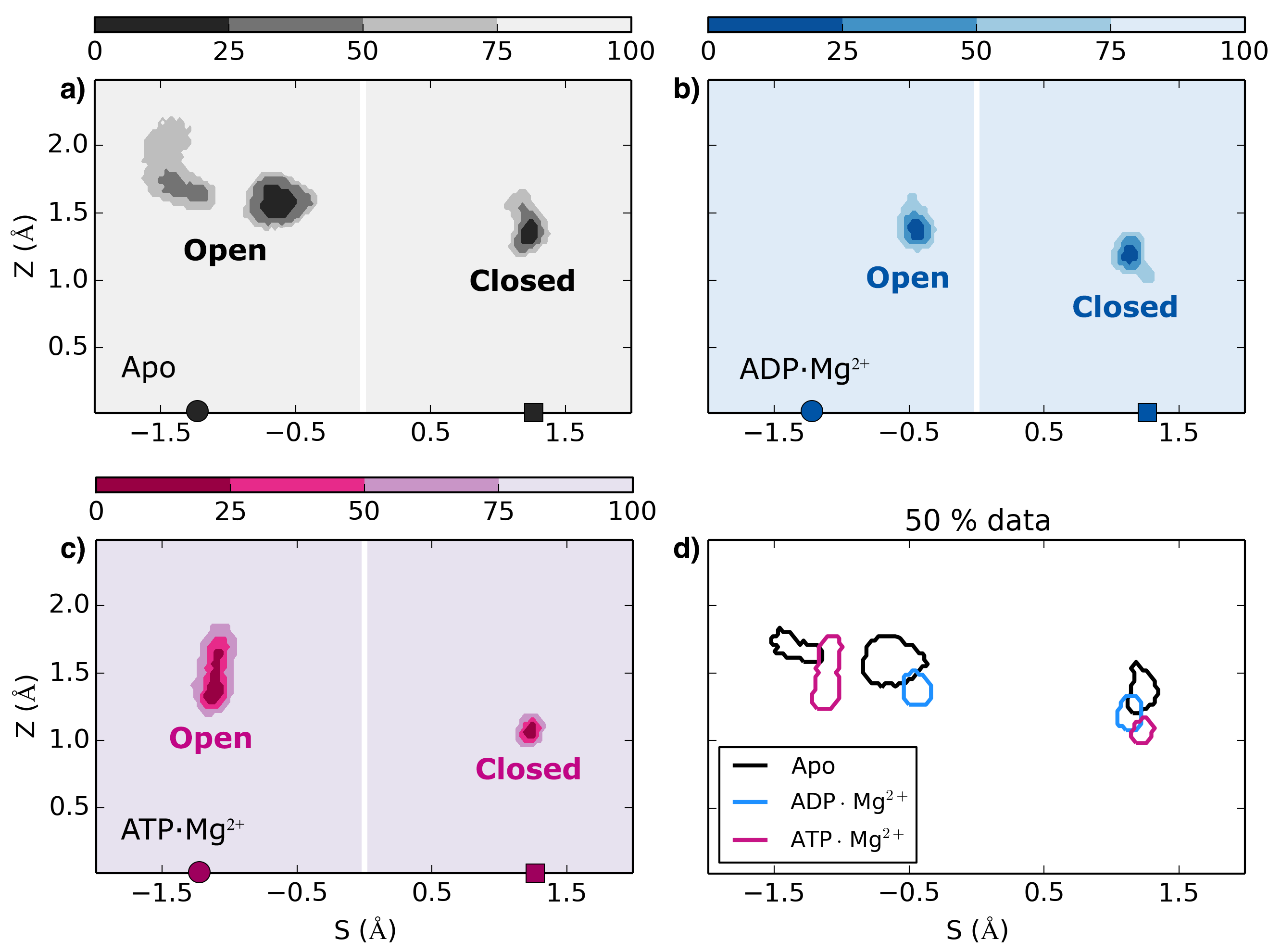}
 \caption{Projection of trajectories on linear path variables
for apo (a), ADP (b), and ATP (c) simulations.
The contour graphs indicate regions containing the indicated fraction of the analyzed conformations
(25, 50, and 75\% as indicated).
Open and closed regions are well defined and do not overlap between them.
Panel (d) summarizes the contour lines corresponding to 50\% of the data.
The location of
open and closed reference structures (PDB: 3O8C, 3O8R) is indicated by a
circle and a square, respectively.}
 \label{fig:sz}
\end{figure}

The linearized path variables $S$ and $Z$ are used here to monitor
conformational changes in our MD trajectories.
As it can be observed in Figure~\ref{fig:sz}, all the simulations 
are stable (see also Figure SD2 and the time series of RMSD in Figure SD3) and remain near to the starting experimental structure ($Z\lesssim 2$\AA)
without undergoing significant transitions to the other conformation.
The protein/RNA complex
is more flexible in the open structure than in the closed one, as it can be appreciated
by the larger fluctuations of the $Z$ value.
Moreover, one can observe that, in presence of the ligand,
fluctuations are partly reduced both in the open and in the closed structure.

The distribution of the $S$ variable indicates that both the apo and the ADP
open systems explore regions that are slightly towards the closed reference structure.
For the apo system a bimodal distribution is observed (Figure~\ref{fig:sz}) with one
peak corresponding to a lower distance from the closed structure.
This peak is however sampled only transiently in the first
half of the simulation, as it can be seen from Figure SD2. In the second half of the trajectory the protein opens again.

We also computed the gap between D1 and D2 as the distance
between the centers of mass of D1 and D2 (see Table SD3).
The interdomain gap is 
$\sim$28 \AA\ for the open conformations and $\sim$26 \AA\ for the closed ones.

Similar results were obtained by the PCA and can be seen in Figure SD4.

\subsection{Hydrogen Bonds}
\begin{table}
\begin{center} 
\caption{Average number of intra-solute hydrogen bonds for each of the six reported simulations.
Detailed counts between selected solute groups are reported in Table SD4-6. }%
\label{tb:hb}                                                                                                                                                                                                                                                                                                                                                                                                                                                                                                                     
\begin{tabular}{|cl|lc|}\hline
\multicolumn{2}{|c|}{\bf Groups} &\bf Total& {$\Delta$HB$_{oc}$}\\\hline\hline
				& Closed				& 349.4	&	\\
{\bf{\color{black}Apo}}&Open				& 348.8	& \ -0.6\\
				&3O8C				& 320	&	\\\hline
\multirow{ 2}{*}{\bf ADP$\cdot$Mg$^{2+}$}&Closed& 353.1&\multirow{ 2}{*}{\ \ 0.5}\\
								&Open& 353.6 & \\\hline
								&Closed& 362.0&	\\
{\bf{\color{black}}ATP$\cdot$Mg$^{2+}$} &Open& 348.5 & -13.5\\
								&3O8R& 329 &	\\\hline
\end{tabular}
\end{center}
\end{table}

We computed the number of hydrogen bonds that are formed during the MD
simulation by several groups of solute atoms, including the three protein domains,
RNA and, when present, ligand.
Detailed results are reported in Supplementary Data Table SD4-6,
whereas the total counts are summarized in Table \ref{tb:hb}.

In general, the values obtained from the simulations starting from the experimental structures
are very similar to those obtained from the crystal structures (columns open-apo vs column 3O8C in Table SD4;
columns closed-ATP vs column 3O8R in Table SD6). However, there are a few discrepancies that must be commented.
In particular, all the counts are slightly higher in the MD trajectories when compared to the crystal structures.
Interestingly, there are a few interactions visible in the MD between domains D1 and D2 for the open structure, 
which are not present in the crystal structures. This is due to a slight closure of the open structure that is
observed during MD and allows contacts to be formed between highly flexible loops.
More precisely, transient hydrogen bonds are formed between an acceptor of the 
Arginine finger (Q460) and donors from the DExH box (E291, H293).
Additionally, hydrogen bonds between domain D1 and D3 are observed in the simulated trajectories of both
the open and closed structures between residues  T305-R512 and S297-E493. These contacts are not present
in the reference crystal structure.
We notice that in both cases the donor-acceptor distances and angles in the crystal structures are 
slightly above the threshold that we used to identify hydrogen bonds and a small fluctuations of these 
bonds increases the hydrogen count in the MD. 
This explanation is valid also for the increased number of intradomain D1-D1 and D2-D2 hydrogen bonds.

Our simulations can also provide an insight on possible intermediate structures that
have not been crystallized, namely closed-apo/ADP and open-ADP/ATP.
We first compare the open and closed structure in the absence of the ligand.
These two structures present a very similar number of hydrogen bonds.
Although the total number of hydrogen bond is not a rigorous estimator
of structural stability, it is expected to give a large contribution to the
interaction energy. Thus, this result suggests that the open and the closed structures
might have a comparable stability in the absence of the ligand.
Notably, the number of D1-D2 hydrogen bonds that are missing in the open structure are compensated
by an equivalent number of intradomain D2-D2 bonds (see Table SD4). At the same time, protein-RNA bonds are overall 
maintained during the simulation, and some D2-RNA bonds are replaced with D3-RNA bonds upon opening.
Very similar considerations can be made for the ADP case (see Table SD5).
On the contrary, when ATP is present the closed structure forms approximately 15 hydrogen bonds
more than the open one. This is due to a combination of several effects (see Table SD6), including
the formation of D1-D2 contacts, the formation of new D2-D2 hydrogen bonds, and the interaction
of the ligand with D2. A detailed structural analysis of the binding pocket is presented below.
The increased number of hydrogen bonds suggests that ATP stabilizes the closed structure.

\subsection{RNA binding cleft}
\begin{figure}\centering
 \includegraphics[width=\columnwidth]{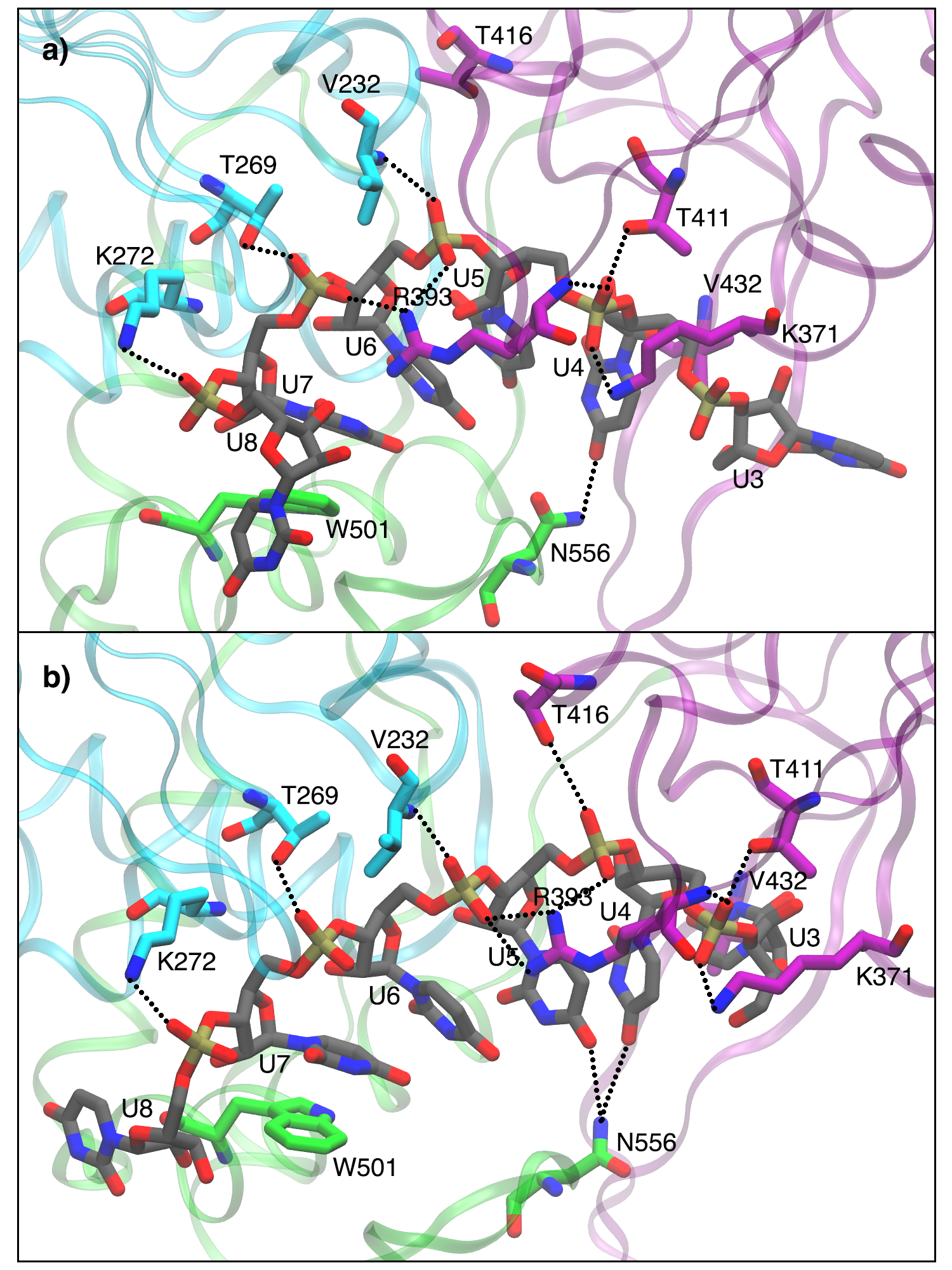}
 \caption{Representation of the RNA binding cleft.
Snapshots depicted correspond to the RMSD centroids of trajectories as obtained by the clustering algorithm
discussed in ref.~\cite{gromos}, using a cutoff radius of 1.25 \AA\ . RNA chain and aminoacids interacting with RNA are highlighted in sticks representation. Residues from D1, D2 and D3 are depicted in cyan, purple and green respectively, RNA chain is shown in gray color. Snapshots are shown for closed-ATP (a) and open-ATP (b) simulations.  Equivalent graphs for the ADP and apo simulations are shown in Figure SD5
}
 \label{fig:rna_cleft}
\end{figure}

The ssRNA interacts with all the three protein domains.
Overall, the contacts between domain D1 and RNA are maintained in the closed and 
open conformations, whereas the contacts between
domain D2 and RNA are either shifted by one nucleotide or missing in the closed structure.
Most of the contacts are with the RNA backbone, consistently with the fact that the helicase can process
RNA irrespectively of its precise sequence. Residues K272, T269 and V232 from domain D1
interact with phosphate oxygens from U8, U7, and U6 respectively in all the six simulations.
In the open structures, T416, T411, and K371 from domain D2 interact with phosphate oxygens from U5, U4, 
and U4 respectively. In the closed structures, interaction of T416 with RNA is not present
and both T411 and K371 interact with phosphate oxygens from U5.
R393 sidechain from domain D2 interacts with phosphates from U6 and U5 in all the open structures. Conversely, in the closed-ATP 
an interaction
with phosphates from U7 and U6  is formed after $\sim$50 ns of simulation.
This interaction is missing in the ADP and apo simulations (see Figure SD5).
Something similar happens to residue N556, which interacts with a 
nucleobase in all the open structures and in the closed-ATP one. 
Finally, we notice that residues W501 (see Figure \ref{fig:rna_cleft}) and V432 (see Figure SD6) 
act as gates located at the 3' terminus and 5' respectively of the simulated oligonucleotide.

\subsection{RNA stacking interactions}
In the closed conformation the RNA strand shows a larger bending so that 
some of the stacking interactions between the nucleobases are lost.
We recall that the 5' terminal base (U3) is lagging outside of the protein in 
the closed structure, and thus is not interacting with the neighbouring base U4.
This stacking is never formed during the MD of the closed structure.
Stacking interactions between U4 and U5 and between U5 and U6 are initially formed 
in the open crystal structure and not formed in the closed crystal structure.
During MD, they appear to be transiently formed also in the latter case, although 
with a lower frequency when compared with the open structures.
Stacking between U6 and U7 appears to be stable in all conditions, since the 
interaction of these nucleotides with D1 and D3 domains
is very similar in the open and closed structures.
Finally, it is important to notice that stacking between the 3' terminal bases U7 
and U8 is never formed. This is due to the stable stacking interaction between U7 and W501 side chain.

\subsection{ATP/ADP binding pocket}
\begin{figure}
\centering
 \includegraphics[width=\columnwidth]{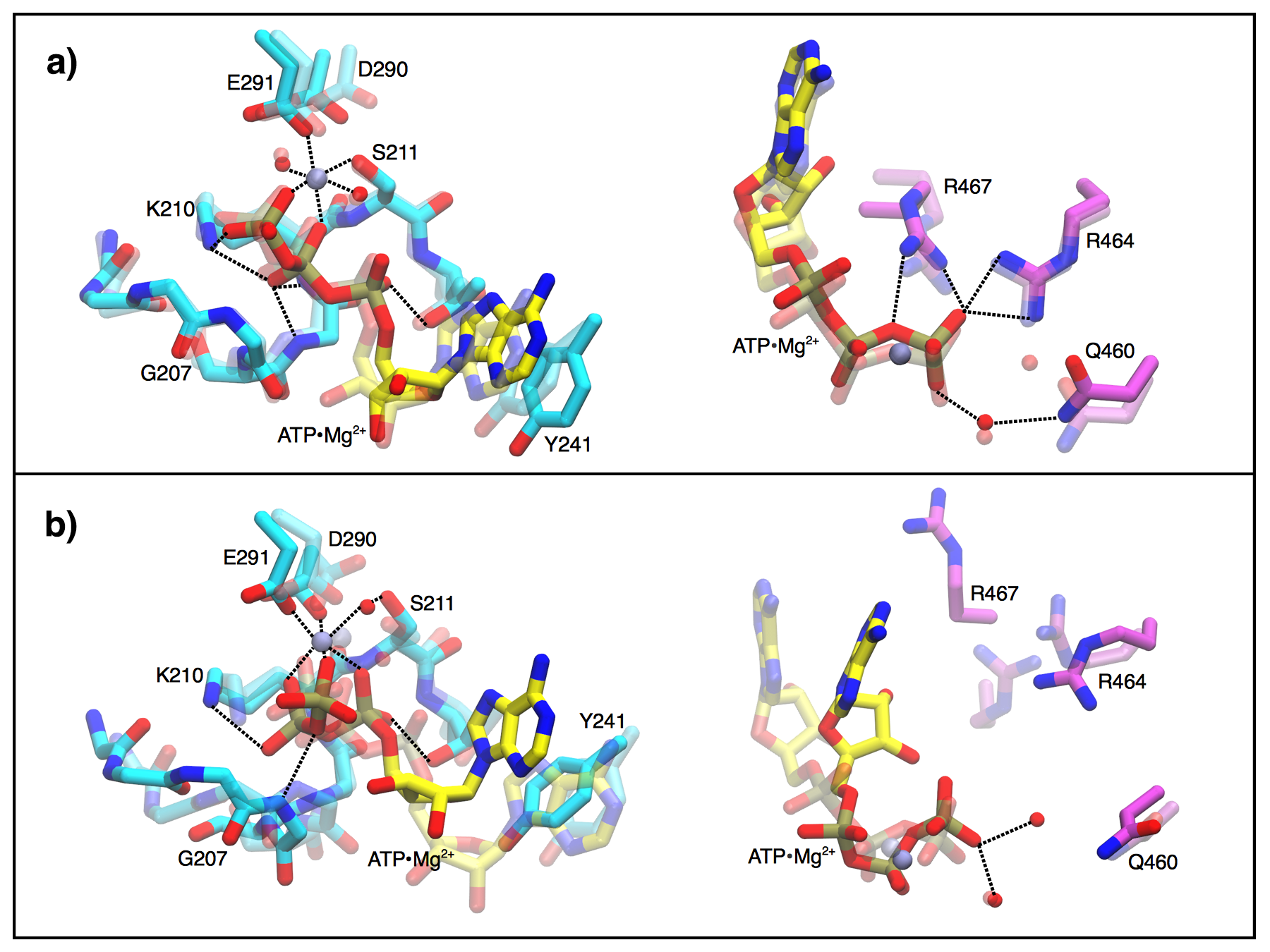}
 \caption{Snapshots of the ATP/ADP binding site.
Aminoacids from D1(cyan) and D2 (purple) and ligand (yellow) are shown as sticks,
the magnesium ion is shown as a lilac sphere.
Closed-ATP (a) and open-ATP (b) systems are shown.
Transparent representation indicates the initial structure,
solid color indicates the structure after 1$\mu$s of molecular dynamics.
Coordination with Mg$^{2+}$ cation and hydrogen bond interactions are shown with the black dashed line.
Equivalent representations for ADP are shown in Figure SD7.
}\label{fig:snaps_atp}
\end{figure}

As expected, the interactions in the binding pocket are heavily 
affected by the protein conformation (open vs closed)
and by the presence and the kind of the ligand (apo vs ADP vs ATP).
As it can be appreciated in Figure \ref{fig:snaps_atp}(a), the closed-ATP 
structure after 1$\mu$s is still very similar to the crystal structure. More precisely, 
interactions between ligand, Walker A (residues 204 to 211),
DExH box (residues 290 to 293), and arginine finger (residues 460 to 467) are preserved
in the MD simulation.
The observed stability of this conformation is consistent with the 
fact that it is among those that have been crystallized~\cite{pyle-jmb}. On the contrary, the open-ATP 
structure experiences a reorganization of the binding pocket after 2 nanoseconds.
In particular, a bending of the ATP is observed which leads to the 
coordination of the Mg$^{2+}$ cation to $\alpha$, $\beta$ and $\gamma$ phosphates, 
which corresponds to an alternative metastable conformation of the ATP/Mg$^{2+}$ complex~\cite{atp_davide}.
Moreover, a contact between Mg$^{2+}$ and D290 (DExH box) is formed 
and the contact between Mg$^{2+}$ and S211 is now mediated by a water molecule.

When ADP replaces ATP (see Figure SD7), both closed and open 
conformations present fluctuations and changes in the interactions between ligand, 
residues and water molecules. Notably, in the open conformation the Mg$^{2+}$ ion 
forms initially a pentacoordinated species which after a few ns is replaced by a 
hexacoordinated species coordinated with 3 water molecules, the $\alpha$ and $\beta$ 
phosphates (Pi's ) and S211. This latter arrangement is stable for the entire simulation.
We observe that pentacoordinated Mg$^{2+}$ is not expected in this system \cite{mespeus} and that 
the force field employed for Mg$^{2+}$ \cite{ff_villamg} can recover the correct hexacoordinated species.

We also notice that whereas in the open-ATP case the ligand exhibits a poor interaction with
domain D2, in the open-ADP case there are several interactions appearing due to a rearrangement 
of R467 that interacts with phosphate $\beta$. This result is also supported by the hydrogen bond analysis in
Table \ref{tb:hb}.
Finally, we observe that in closed-ADP and open-ATP, the stacking between the adenine 
from the ligand and Y241 is formed for a small fraction of the time. This suggests that stability 
of the ligand is enhanced in closed-ATP structure relatively to the other cases.

\subsection{Electrostatic interactions}

Debye-H\"uckel interaction energies (G$^{DH}$) were computed discarding the initial 200 ns of simulation.
These numbers cannot provide quantitative estimates of the binding affinities but can provide a qualitative ranking
for the different protein conformations.

We computed the interaction between the peptide and the RNA so as to have a proxy
for the RNA binding energy (see Table \ref{tb:dhe}). 
The RNA-protein interaction is slightly larger ($\sim$1 kcal/mol) in the closed conformations when compared with the open ones,
irrespectively of the presence of the ligand.
However, we notice that in the presence of ATP the RNA-protein interaction is stronger. By considering separately the direct
RNA-ligand interaction (also shown in Table \ref{tb:dhe}) it can be appreciated that this change is
due to a rearrangement of the protein and not to a direct RNA-ligand electrostatic coupling.
The contribution of the individual aminoacids located at distance $<$6 \AA\ from the RNA is  presented in Table SD7
where it can be appreciated that in the closed conformations the most interacting residue is K371 whereas for the open structures
the most interacting residue is R393. An exception is given by the closed-ATP where also the R393 is the most interacting residue.
This difference is consistent with the rearrangement of the RNA binding cleft discussed above.

We also computed the interaction between the ligand and the RNA-protein complex, which is also reported in Table \ref{tb:dhe}.
Remarkably, ATP interacts more with the protein in its closed conformation, whereas ADP interacts more with the protein in its open conformation.
Though being far from quantitative, this result is qualitatively consistent with the fact that the affinity for ATP is larger in the closed form.

\begin{table}[!h]
\begin{center}
\caption{Electrostatic interaction computed as Debye-H\"uckel energies (G$^{DH}$). Protein* denotes Protein-RNA complex. Errors were computed from binning analysis (bin width: 80ns). Errors lower that 0.05 kcal/mol are shown as "0.0".} %
\label{tb:dhe}
\begin{tabular}{|c|rr|}\hline
			 & \multicolumn{2}{c|}{\bf $\Delta$G$^{DH}$ (kcal/mol)}   \\ %
 \bf{RNA-Protein}& \bf Closed & \bf Open       \\ \hline%
 {\color{black}Apo}		& -4.9 $\pm$ 0.0 & -3.7 $\pm$ 0.2 \\ 
 {ADP$\cdot$Mg$^{2+}$}	& -5.0 $\pm$ 0.1 & -3.9 $\pm$ 0.0 \\
 {ATP$\cdot$Mg$^{2+}$}	& -5.5 $\pm$ 0.1 & -4.7 $\pm$ 0.1 \\\hline\hline
\bf RNA-Ligand & \bf Closed & \bf Open      \\ \hline%
ADP$\cdot$Mg$^{2+}$	& -0.1 $\pm$ 0.1 & -0.2 $\pm$ 0.0 \\
ATP$\cdot$Mg$^{2+}$	& -0.2 $\pm$ 0.1 & -0.2 $\pm$ 0.1 \\\hline\hline
\bf Ligand-Protein* & \bf Closed & \bf Open      \\ \hline%
ADP$\cdot$Mg$^{2+}$	& -0.4 $\pm$ 0.0 & -1.1 $\pm$ 0.0 \\
ATP$\cdot$Mg$^{2+}$	& -0.8 $\pm$ 0.0 & -0.1 $\pm$ 0.0 \\\hline
\end{tabular}
\end{center}
\end{table}

\subsection{Enthalpies}
Average enthalpies were computed discarding the initial 200 ns of each trajectory.
This initial part is thus considered here as equilibration.
We only computed enthalpy differences between simulations
including exactly the same number and types of atoms,
corresponding to two alternative conformations (open vs closed).
From the results reported in Table \ref{tb:enthalpy}, 
it is observed that open conformation has a systematically lower enthalpy than closed ones.
This difference is more marked when ATP is present. We observe that statistical errors, as obtained
from binning analysis (bin width: 80 ns), are very low for all the cases.
Estimation of the enthalpy values for the first and second halves of the simulations are also reported in 
Table SD8 and the values are consistent with the reported ones from the entire trajectory.

\begin{table}[!h]
\begin{center}
\caption{Enthalpy differences between open and closed conformations. The values are computed between systems with same number and kind of molecules.} %
\label{tb:enthalpy}
\begin{tabular}{|c|r|}\hline
\bf{Ligand} & \bf{$\Delta H_{oc}$\ \ \ \ }   \\
           & \bf{(kcal/mol)}        \\ \hline\hline
 \bf{\color{black}Apo}		&  -9.7 $\pm$ 2.8 \\ 
 \bf{\color{black}ADP$\cdot$Mg$^{2+}$}	& -25.3 $\pm$ 3.1 \\
 \bf{\color{black}ATP$\cdot$Mg$^{2+}$}	& -54.1 $\pm$ 4.1 \\\hline
\end{tabular}
\end{center}
\end{table}

Enthalpy values are important to describe internal interactions \cite{freire1996}.
Differences of enthalpy thus provide an indication of relative stability
of different conformations, at least when their contribution is dominant in the
free-energy difference.
In the present context, enthalpy calculations suggest that the ATP stabilizes the
open structure,
in contrast with the observed number of intra-solute hydrogen bonds, the structural analysis of the binding pocket,
and the electrostatic interactions estimated within the Debye-H\"uckel model.

\subsection{Targeted molecular dynamics}
The work performed during the targeted molecular dynamics was used as a proxy for the 
relative stability of the closed and open structure in the presence and the absence of the ligand. One should consider here
that the average work only provides an upper estimate of the free-energy change, and that
in principle simulations should be combined using the Jarzynski's equality \cite{jarzynski}.
However, a very large number of simulations would be required to provide a converged estimate of the free-energy change
for this complex rearrangement.
Additionally, since in the targeted MD an increasingly strong restraint is applied to a large number of atoms, the absolute value of the free-energy
change would be largely affected by the decrease in the entropy of the restrained atoms.
In spite of these quantitative limitations, the average work can be used as a qualitative tool to rank the free-energy differences
for apo, ADP, and ATP structures.
The results for the different steering protocols are shown in Figure SD8.
Here, it can be appreciated that the work required to open the apo form is systematically lower than the one required
to open the ADP and ATP forms. This result is
compatible with the structural analysis of the binding pocket, the hydrogen bonds between solute molecules, and
the electrostatic interactions discussed above,
indicating that the enthalpy is compensated by a large entropic change.
 The ligand-induced stabilization of the
closed form is also compatible with the fact that the ATP form was crystallized in this conformation. 
\section{Discussion}
We here presented a detailed analysis of conformational properties of NS3 helicase from HCV in complex with ssRNA,
with ATP/ADP, and in the apo form. For each of these three systems, we performed MD simulations starting from both
the open and the closed conformation, for a total of 6 simulations of 1 $\mu$s length each.
For two of these simulations the starting structure was already available as obtained from X-ray cristallography,
whereas the remaining four systems were built by structural alignment.
Although this protein/RNA complex has only been crystallized either in the apo form or with ATP, we 
also analyzed the effect of ATP replacement by ADP.
An intermediate structure with ADP has been reported for NS3 from 
Dengue virus \cite{dengue}, but has never been characterized for the HCV NS3.

The ATP hydrolysis reaction was not explicitly analyzed here, but has
been addressed in the literature for the PcrA helicase \cite{pcra} and for other motor proteins (see e.g. \cite{h-loop, f1-atpase, kinesin}).
The present work is not aimed at 
giving a mechanistic understanding of hydrolysis itself,
but rather at providing an insight on the system's behavior after the hydrolysis reaction
has occurred.

Overall, all the observed trajectories were stable on this timescale and did not experience any significative structural change.
This provides a validation for the protocol employed to build the four structures not obtained from X-ray.
In particular, structural stability was assessed by monitoring combinations of RMSD from the open and closed PDB structures,
where it can be appreciated that all systems remained near to the experimental structure on this timescale.
Interestingly, the ATP and ADP open structures have a lower enthalpy than the corresponding closed structures,
indicating that they have been properly equilibrated.
Additionally, we observed that in the presence of ATP the closed structure fluctuated less compared with closed-apo/ADP 
and its open analogue, %
suggesting that the closed comformation is stabilized by the presence of ATP.
This is consistent with
the fact that the crystallized structure with ATP is in its closed conformation \cite{pyle-jmb, gu_rice}.
On the other hand, the simulations from the open structure displayed larger fluctuations.
This holds also for the open-apo structure, which has been crystallized \cite{pyle-jmb, gu_rice}.
It is possible that crystal contact or interactions with the protease domain
stabilize the open-apo structure in the experiments.

When ATP is present, the total number of hydrogen bonds formed by the solute is significantly larger in the closed
structure when compared with the open one. This is also consistent with the fact that ATP stabilizes the closed
conformation. Interestingly, for both the apo and the ADP complex this is not observed, and the number of hydrogen 
bonds is equivalent in the open and closed structures.
This indicates that the free-energy difference $\Delta G$ between the open and the closed structures
should be larger in the ATP case, as confirmed also by the targeted MD simulations.
This $\Delta G$
is related to the differential affinity of the ligand in the two structures. Indeed,
$\Delta G_{ATP}-\Delta G_{Apo} = -k_BT\log K_d^{(closed)} + k_BT\log K_d^{(open)}$, where $k_{B}$ is the Boltzmann constant,
$T$ is the temperature, and $K_d^{(open)}$ and $K_d^{(closed)}$ are the ATP dissociation constants in the open and closed
conformation respectively. Thus, our result implies that the ATP affinity is larger in the closed form.
The difference in the protein-ADP/ATP interaction estimated by a Debye-H\"uckel model is also consistent with this picture.

Surprisingly, enthalpy calculations has
an opposite trend, suggesting that open-ATP is dramatically more stable than closed-ATP.
We observe that this difference ($\approx 54$kcal/mol) is much larger than its statistical error ($\approx 4$kcal/mol)
so that the reported MD simulations are sufficiently converged to estimate this trend.
We recall that, although enthalpy provides a very important contribution to free-energy changes,
it might be not sufficient to evaluate the relative stability of two conformers. Indeed,
entropic contributions have been reported to cancel enthalpic ones in several cases~\cite{eec_chodera}. 
From our simulations, it is not possible to properly estimate the entropic difference which should be
ascribed not only to the changes in solute flexibility but also to non trivial changes in water entropy which
are connected with e.g. hydrophobic effects in the cavity between D1 and D2. 
Although it is in principle possible to use enhanced sampling
methods to directly compute free-energy differences~\cite{dellago2014computing,abrams2014enhanced}, this is a complex procedure
for concerted conformational changes and is
left as a subject of further investigation. From the presented results, it can be inferred that a strong
entropic compensation stabilizes the experimentally observed closed-ATP structure.

We also analyzed carefully the pattern of RNA-protein interactions.
In our simulations the protein does not interact with sugars in the RNA backbone. This is consistent 
with the empirical observation that this helicase can also process DNA \cite{myong, gu_rice}.
Contacts between T269 and T411 with the phosphate oxygens are different according to the protein conformation 
and act like hooks with the RNA chain during the translocation mechanism.
This feature has been reported for other SF2 RNA-binding helicases such as Vasa drosophila and elF4A,
suggesting a possible common mechanism among several helicases of this superfamily \cite{myong}.
Besides confirming the contacts that were already seen in the experimental structures,
our simulations provide an insight on the possible conformations available for the open-ATP and closed-apo structures.
Indeed, in the ATP complex  we observe different contacts with residues R393 and N556, which should be likely ascribed
to an allosteric change induced by the ligand.

We notice that the contact network of the ligand binding pocket is affected to the largest extent by the addition of ATP/ADP or by opening and closing.
The closed-ATP complex preserves the contacts between the Motifs I, VI and the ATP observed in the crystal structures.
On the contrary, the open-ATP complex presents a rearrangement due to the missing contacts with D2.
The above effect is probably an underlying reason for the ATP-dependent stabilization of the closed form.
Upon the replacement of the ATP by ADP, which mimics the result of the hydrolysis reaction, more interactions are observed 
between the ligand and D2. This suggests that the stabilizing effect of ADP on the open form is significantly larger than that of ATP. 
In other words, after hydrolysis has occurred the closed conformation is destabilized with respect to the open one and a conformational transition towards to open structure could be favored.

As a final remark, we observe that performing molecular dynamics of RNA/protein complexes is a non trivial task,
and only a few studies have been published so far (see Ref. \cite{krepl2015can} for a recent overview).
In this work, we tested molecular dynamics on a biologically relevant complex including a medium sized protein, a short ssRNA,
and a nucleotide ligand using recent force fields
on the $\mu$s time scale, which is the state of the art for classical molecular dynamics.
The total  time including all the systems and all the control simulations is approximately 9 $\mu s$.
This simulation time allowed us to compute averages with relatively small statistical errors
 and to perform several consistency checks indicating that our simulations
 provide a converged conformational ensemble around the equilibrium structures.
The experimental snapshots are stable within the investigated time scale, 
indicating that these force fields might be able to properly characterize such heterogenous complexes.
Although enhanced sampling techniques should be used to quantify the relative stability of the simulated conformations,
this work provides an extensive characterization of the possible intermediates sampled during NS3 translocation on RNA
at atomistic details.

\section{Funding}
The research leading to these results been funded by the European Research Council under the European Union's Seventh Framework Programme (FP/2007-2013) / ERC Grant Agreement n. 306662, S-RNA-S. 
We acknowledge the CINECA award no.~HP10BL6XFZ, 2013, under the ISCRA initiative for the availability of high performance computing resources.

\subsubsection{Conflict of interest statement.} None declared.

\section{Acknowledgements}
Anna Marie Pyle is acknowledged for useful discussions
and for providing enlightening suggestions.
Petr Stadlbauer and Jiri Sponer are also acknowledged for useful discussions.

\bibliographystyle{nar}

\end{document}